\documentclass[fleqn,10pt]{wlscirep}
\usepackage[utf8]{inputenc}
\usepackage[T1]{fontenc}
\usepackage{float}
\title{Diversity enables the jump towards cooperation for the Traveler's Dilemma}

\author[1,2,*]{Mar\'{i}a Alejandra Ram\'{i}rez}
\author[2]{Matteo Smerlak}
\author[1]{Arne Traulsen}
\author[2,3]{J\"{u}rgen Jost}
\affil[1]{Max Planck Institute for Evolutionary Biology, Pl\"{o}n, 24306, Germany}
\affil[2]{Max Planck Institute for Mathematics in the Sciences, Leipzig, 04103, Germany}
\affil[3]{Santa Fe Institute for the Sciences of Complexity, Santa Fe, NM, 87501, USA}

\affil[*]{ramirez@mis.mpg.de}


\begin{abstract}
Social dilemmas are situations in which collective welfare is at odds with individual gain. One widely studied example, due to the conflict it poses between human behaviour and game theoretic reasoning, is the Traveler's Dilemma. The dilemma relies on the players' incentive to undercut their opponent at the expense of losing a collective high payoff. Such individual incentive leads players to a systematic mutual undercutting until the lowest possible payoff is reached, which is the game's unique Nash equilibrium. However, if players were satisfied with a high payoff -that is not necessarily higher than their opponent's- they would both be better off individually and collectively. Here, we explain how it is possible to converge to this cooperative high payoff equilibrium. Our analysis focuses on decomposing the dilemma into a local and a global game. We show that players need to escape the local maximisation and jump to the global game, in order to reach the cooperative equilibrium. Using a dynamic approach, based on evolutionary game theory and learning theory models, we find that diversity, understood as the presence of suboptimal strategies, is the general mechanism that enables the jump towards cooperation.
\end{abstract}
\begin{document}

\flushbottom
\maketitle

\thispagestyle{empty}

\section*{Introduction}
Game theory is a framework for analysing the outcome of the strategic
interaction between decision makers \cite{fudenberg:book:1998a}. The fundamental concept is that of a Nash equilibrium where no player can improve her payoff by a unilateral strategy change. Typically, the Nash equilibrium is considered to be the optimal outcome of a game, however in social dilemmas the individual optimal outcome is at odds with the collective optimal outcome \cite{broom:BMB:2018}. This means that one player can improve her payoff at the expense of the other by unilaterally deviating, but if both deviate, they end up with lower payoffs. In this type of games, the mutually beneficial, but non-Nash equilibrium strategy is called cooperation. However, in this context cooperation should not be interpreted as altruism, as players only aim to secure a high payoff for themselves.\\

In this framework, payoff maximisation is considered to be rational, but when such rational players then seize every opportunity to gain at the opponent's expense, they
may counterintuitively both end up with low payoffs. A game that clearly exhibits this contradiction is the Traveler's Dilemma. Since its formulation in 1994, by the economist Kaushik Basu \cite{basu:AEA:1994}, the game has become one of the most studied in the economics literature. Additionally, it has been discussed in theoretical biology in the context of evolutionary game theory.\\

In general, the dilemma relies on the individuals' incentive to undercut the opponent. To be specific, players are motivated to claim a lower value than their opponent to reach a higher payoff than them. Such incentive leads players to a systematic mutual undercutting until the lowest possible payoff is reached, which is the unique Nash equilibrium. It seems paradoxical that players defined as rational in a game theoretical sense end up with such a poor outcome. Therefore, the question that naturally arises is how can this poor outcome be prevented and how cooperation can be achieved.\\

To address these questions can be helpful to better understand price wars, which consist on the mutual undercutting of prices to gain market share.
In addition, it can provide information about human behaviour. As economic experiments have shown that individuals prefer to choose the cooperative high payoff action, instead of the Nash equilibrium \cite{capra:AER:1999}.\\

Our analysis focuses on showing that the Traveler's Dilemma can be decomposed into a local and a global game. If the payoff optimisation is constrained to the local game, then players will inevitably end up in the Nash equilibrium. However, if players escape the local maximisation and optimise their payoff for the global game, they can reach the cooperative high payoff equilibrium.\\

Here, we show that the cooperative equilibrium can be reached in a game like the Traveler's Dilemma thanks to diversity, which we define as the presence of suboptimal strategies. The appearance of strategies far from those of the residents allow for the local maximisation process to be escaped, such that an optimisation at a global level takes place. Overall this can lead to cooperation because by considering ``suboptimal strategies" that play against each other it is possible to reach higher payoffs, both collectively and individually.

\subsection*{Game} \label{game}
The Traveler's Dilemma is a two-player game. Player $i$ has to choose a claim, $n_i$, from the action space, consisting of all integers on the interval $[L,U]$, where $0 \leq L < U$. The payoffs are determined as follows:
\begin{itemize}
    \item If both players, $i$ and $j$, choose the same value ($n_i = n_j$),
      both get paid that value.
    \item There is a reward parameter $R>1$, such that if $n_i < n_j$, then $i$
      receives $n_i + R$ and  $j$ gets  $n_i- R$
\end{itemize}

Thus, the payoff of player $i$ playing against player $j$ is

\begin{equation}
\label{eq:payoff}
\pi_{ij} =
\begin{cases}
n_i \text{ \hspace{10mm} if } n_i = n_j\\
n_i + R \text{ \hspace{3mm} if } n_i < n_j\\
n_j - R \text{ \hspace{3mm} if } n_i > n_j
\end{cases}
\end{equation}

Thus, a player is better off by choosing a slightly lower value than its opponent: when $j$ plays $n_j$, then it is best for $i$ to play $n_j-1$. The iteration of this
reasoning, which we will call the \textit{stairway to hell}, leads to the only Nash equilibrium of the game, $\{L,L\}$, where both players choose the lowest possible claim. The classical game theory method to arrive to this equilibrium is called iterative elimination of dominated strategies \cite{hofbauer:book:1998}.\\

\begin{figure}[ht]
\centering
\includegraphics[width=12cm]{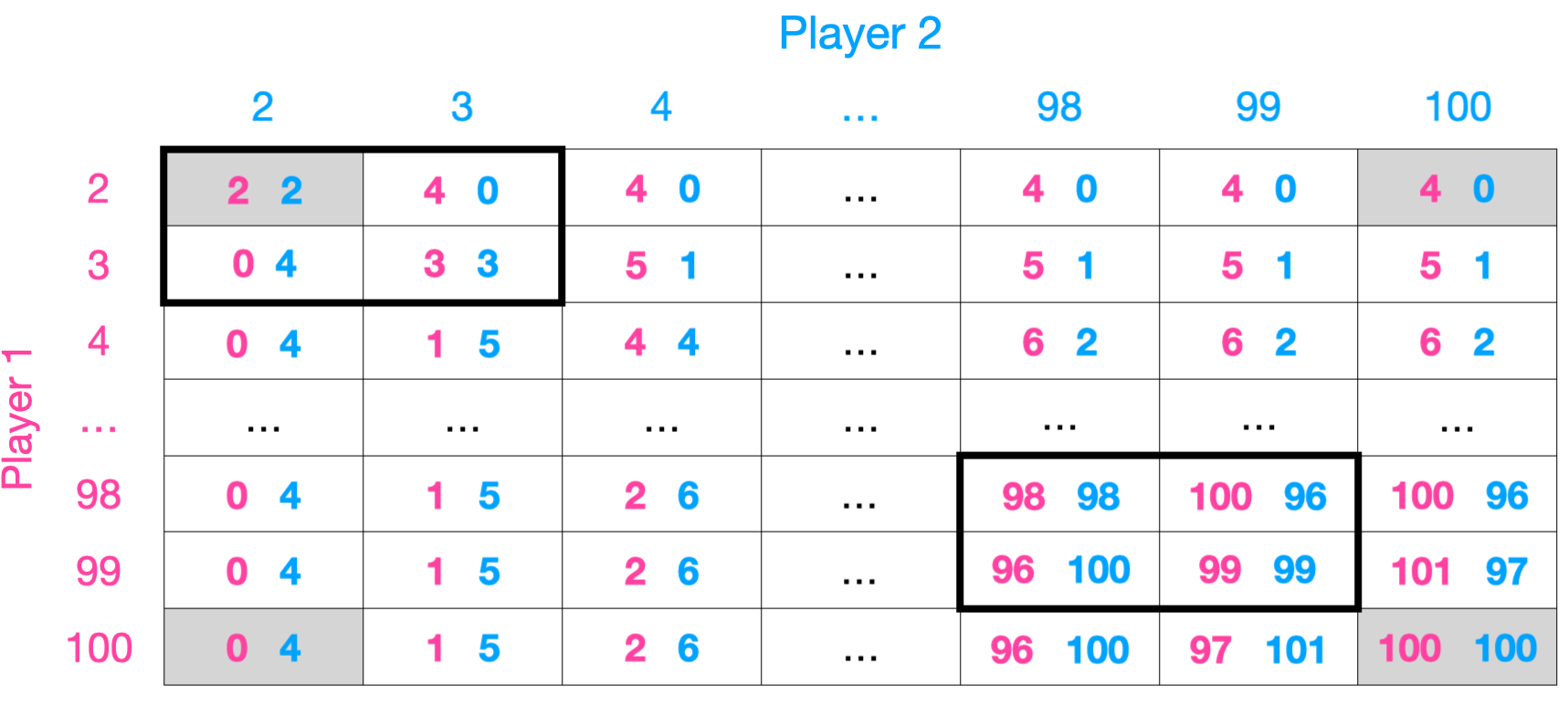}
\caption{\textbf{Payoff matrix of the Traveler's Dilemma.} Visualisation of the payoff scheme described by equation \ref{eq:payoff}. For simplicity, the action space is $ \{n_i \in \mathbb{N} \mid 2 \leq n_i \leq 100\}$ and the reward parameter is $R=2$.\\ The Traveler's Dilemma can be decomposed into a local and a global game. At a local level the game is a Prisoner's Dilemma (black boxes). This happens when the action space is reduced to any pair of actions $n$ and $n+s$, where $\{s \in \mathbb{N} \mid 1 \leq s \leq R-1 \}$. While at the global level, we can observe a coordination game (gray boxes). This level is defined as any pair of actions $n$ and $n+s$, where $\{s \in \mathbb{N} \mid s \geq R \}$}
\label{fig:Payoff_M}
\end{figure}

The game can be visualised through its payoff matrix (Fig \ref{fig:Payoff_M}). For simplicity, we use the values from the original formulation: $L=2$, $U=100$ and $R=2$.
The payoff matrix shows that the Traveler's Dilemma can be decomposed into a local and a global game. Let us begin with the local game. When the action space of the game is reduced to two adjacent actions $n$ and $n+1$ (black boxes in Fig \ref{fig:Payoff_M}), the Traveler's Dilemma with $R=2$ is equivalent to the Prisoner's Dilemma \cite{rapoport:book:1965}. In general, for any value of $R$, the Traveler's Dilemma becomes a Prisoner's Dilemma for any pair of actions $n$ and $n+s$, where $\{s \in \mathbb{N} \mid 1 \leq s \leq R-1 \}$. For example, for $R=4$ the pair of actions $n$ and $n+1$, $n$ and $n+2$, $n$ and $n+3$ follow the same game structure as the Prisoner's Dilemma. Therefore, the Traveler's Dilemma consists of many embedded Prisoners' Dilemma. This means that at a local level the game is a Prisoner's Dilemma.\\

If we now consider actions that are distant from each other in the action space, e.g. $2$ and $100$, we can observe a coordination game structure (gray boxes in Fig \ref{fig:Payoff_M}); where $\{100,100\}$ is payoff and risk dominant \cite{fudenberg:book:1998b, bayer:PLosOne:2022}. In general, any pair of actions $n$ and $n+s$, where $\{s \in \mathbb{N} \mid s \geq R \}$, construct a coordination game. As a result, the Traveler's Dilemma becomes a coordination game at a global level, which has a different equilibrium from the local game.\\

\subsection*{Paradox}
Social dilemmas appear paradoxical in the sense that self-interested competing players, when rationally playing the Nash equilibrium, end up with a payoff that clearly goes against their self-interest. But with the Traveler's Dilemma, the paradox goes further, as it has been suggested in its original formulation \cite{basu:AEA:1994}.
Classical game theory proposes $\{L,L\}$ as the Nash equilibrium of the game. However, it seems unlikely and implausible that, with $R$ being moderately low, say $R=2$, for individuals to play the Nash equilibrium. This has been confirmed in economic experiments where individuals rather choose values close to the upper bound of the interval. Such experiments have also shown that the chosen value depends on the reward parameter (R), where an increasing value of R shifts players' decision towards $\{L,L\}$ \cite{capra:AER:1999}. Nonetheless, classical game theory states that the Nash equilibrium of the game is independent of R.\\

Consequently, the aim of this paper is to seek and explore simple mechanisms through which the apparent non-rational cooperative behaviour can come about.
We also examine the effect of the reward parameter on the game outcome.
Given that the Traveler's Dilemma paradox emerges in the classical game theory framework, we analyse the game using evolutionary game theory tools \cite{maynard-smith:Nature:1973, weibull:book:1995, hofbauer:book:1998}. This dynamical approach allows us to explore adaptive behaviour outside of the stationary classical game theory framework. To be more precise, for this approach individuals dynamically adjust their actions over time according to their payoffs. The game dynamics can then lead to an optimal outcome without invoking constrained assumptions on the players' rationality.\\

The key point of course is to understand how the system can converge to high claims.
We show that this behaviour is possible because the Traveler's Dilemma can be decomposed into a local and a global game. If the payoff maximisation is constrained to the local level, then the \textit{stairway to hell} leads the system to the Nash equilibrium; given that locally the game is a Prisoner's Dilemma. On the other hand, at a global level the game follows a coordination game structure, where the high claim actions are payoff dominant. Thus, for the system to reach a high claim equilibrium the maximisation process needs to jump from the local to the global level.\\

Our analysis led us to identify the mechanism of \emph{diversity} as the responsible for enabling this jump and preventing players from going down the \textit{stairway to hell}. This mechanism works on the idea that to reach a high claim equilibrium, players have to benefit from playing a high claim. For a population setting, it means that players need to have the chance to encounter opponents also playing high. From a learning model point of view, it refers to the belief that the opponent will also play high, at least with a certain probability. If the belief is shared by both players, they should both play high and reach the cooperative equilibrium.
Here, we explore these two types of models that unveil the mechanism leading to cooperation.\\

Population based models unveil diversity as the cooperative mechanism via the effect of mutations on the game's outcome. As it is shown for the replicator-mutator equation and the Wright-Fisher model.
Alternatively, a two-player learning model approach, more in line with human reasoning, shows that if players are free to adopt a higher payoff action from a diverse action set during their introspection process, they can reach the cooperative equilibrium. This result is obtained using introspection dynamics.\\

Finally, we explain how diversity is the underlying mechanism
that enables the convergence to high claims in previously proposed models. We show that diversity is required because it allows for the maximisation process to jump from the local to the global level.

\section*{Replicator-Mutator Equation}\label{RepMutEq}
\subsection*{Formulation}
The replicator-mutator equation is a generalization of the replicator dynamics, which is a fundamental model to describe selection in a population. As the replicator dynamics, the equation describes the evolution of an infinite population of $n$ different types whose frequencies are $x_1,...,x_n$. To model selection, the reproduction rate of each type $i$ is determined by its fitness $f_i$, which is derived from the payoff matrix of interactions among individuals in the population \cite{roca:PLR:2009}. In addition to the typical formulation of the replicator dynamics, the equation includes mutations via the mutation matrix, $Q = (q_{ji})$, where $q_{ji}$ is the probability that type $j$ mutates to type $i$ \cite{broom:book:2013, bauer:PRSA:2019}. Consequently, the mutation matrix $Q$ is a row-stochasic matrix. Thus the replicator-mutator equation is described as follows:

\begin{equation}
\label{eq:rep_mu_eq}
\dot{x_i} = \sum^{n}_{j=1}x_jf_j(\vec{x})q_{ji} - x_i\phi \hspace{5mm} i=1,..,n
\end{equation}

where $\vec{x}=(x_1,x_2,...,x_n)$ and $\phi=\sum^{n}_{i=1}x_if_i(\vec{x})$ denotes the average population fitness.\\
To model mutations, we consider a uniform random probability of mutating to other types, i.e.

\begin{equation}
\label{eq:Mutations}
q_{ij}=\frac{q}{n-1},\hspace{2mm} i \neq j, \hspace{5mm} q_{ii} = 1-q,  \hspace{2mm} 1\leq i, j \leq n
\end{equation}

The strength of mutation parameter $q$ can take values between $0$ to $\frac{n-1}{n}$. When $q=0$, the dynamics corresponds to the replicator dynamics, such that no mutations occur. While for $q=\frac{n-1}{n}$, type $i$ is equally likely to change to any other type or to remain as it is \cite{duong:DGAA:2020}.

\subsection*{Results}\label{RepMutEqresults}
\begin{figure}[ht]
\centering
\includegraphics[width=17.5cm]{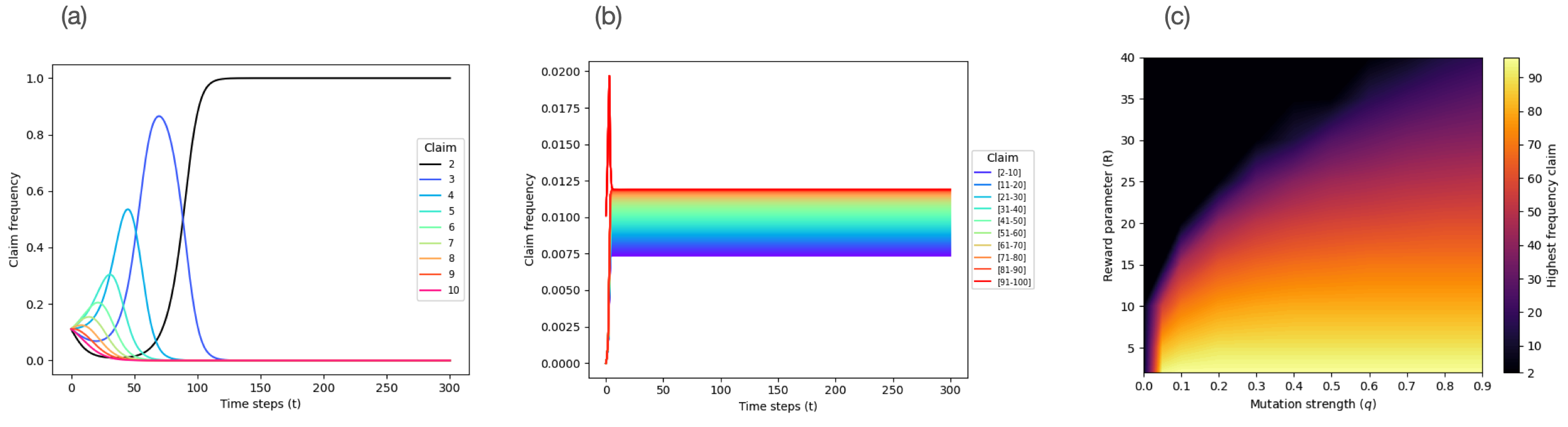}
\caption{\textbf{Replicator-mutator equation for the Traveler's Dilemma.} \textbf{a)} Replicator dynamics limit ($q=0$). The dynamics show that dominated actions are systematically eliminated from the population. Only $L$ remains and dominates the population. Without loss of generality, here the interval $[2,10]$ is considered to clearly visualise the systematic elimination of actions on the same time scale. \textbf{b)} Replicator-mutator dynamics for $q=0.7$ and $R=2$. After the initial transient, the system converges to a coexistence equilibrium with a highest frequency claim, which in this case is 96. \textbf{c)} Highest frequency claim reached by the replicator-mutation dynamics for different reward parameters (R) and mutation strength ($q$) values. The results are in agreement with economic experiments, where high claims are reached with low reward values. Overall, the numerical solution reveals that mutations are necessary for high claims to be benefitted in the population. This means that diversity should be maintained in the population for high claims to be favourable.}
\label{fig:RepMutGraph}
\end{figure}

The replicator dynamics limit ($q=0$) results are in agreement with classical game theory predictions. As expected, $L$ is the claim that dominates the population regardless of the reward parameter value (R).
Specifically, for this dynamics the \textit{stairway to hell} takes place, where dominated strategies are systematically eliminated through extinction until only the Nash equilibrium claim remains in the population (Fig \ref{fig:RepMutGraph}.a). This behaviour is related to the theorem proposed in Hofbauer \& Sandholm. (2011) \cite{hofbauer:TECO:2011}, where dynamics -like the replicator dynamics- successfully eliminate strictly dominated strategies, given that they fulfil a set of requirements.\\

When mutations are allowed ($q\neq0$), a different behaviour can be observed. The system reaches a coexistence equilibrium, where there is a highest frequency claim (Fig \ref{fig:RepMutGraph}.b). No extinction or dominance of actions is present, due to the sustained presence of mutation events. In figure \ref{fig:RepMutGraph}.c, the results for different reward parameters (R) and mutation strength ($q$) values are shown. The numerical solution reveals that mutations are necessary for high claims to be benefitted in the population. This means that diversity should be maintained in the population for high claims to be favourable. Also, the results exhibit the same behaviour than economic experiments in terms of the reward parameter, where a low R value is needed for high claims to be played most frequently.

\section*{ Wright-Fisher Model}\label{WF}
\subsection*{Model}

The basic model for the evolution of populations including
genetic drift, mutation and selection is the Wright-Fisher model, see
e.g. \cite{imhof:JMB:2006, otto:book:2007, etheridge:book:2011, hofrichter:book:2017}. It is therefore natural to also analyse the Traveler's Dilemma within that framework. In particular, our aim is to further explore how the population can converge to high claims.\\

The Wright-Fisher model describes the evolution of a population with
discrete, non-overlapping generations of fixed size $N$. From the current
generation, the next generation is obtained by sampling with replacement, such that the sampling weights are proportional to the fitness of the individuals. For each sampled individual there is a probability $\mu$ of mutation.\\

In our context, an individual is a player of the Traveler's Dilemma with a
certain value as her claim, such that a mutation means that the value of the claim
changes by $\pm \delta$; unless this would result in a value smaller than $L$
or larger than $U$, in which case simply $L$ or $U$, respectively, is chosen.\\

The expected payoff of each individual is determined by playing the Traveler's Dilemma with the other individuals present in the population, excluding self-interactions \cite{datseris:Simulation:2022}.
The accumulated payoff $\Pi$ then defines the fitness of an individual, that is, the sampling weight in the replacement scheme, via
\begin{equation}
\label{eq:WF_Fitness}
f = e^{\rho\Pi}
\end{equation}
where the selection intensity coefficient $\rho$ modulates the magnitude of the impact of fitness differences in the game dynamics. The larger the selection intensity, the stronger the influence of fitness differences and the less important is the role of stochasticity on the dynamics.\\

The process of fitness dependent sampling with replacement and mutation
probability $\mu$ is repeated for $t$ time steps. We fix the population size to $N=100$ and consider $\mu, \delta, \rho$ as variable parameters. We end the simulation after $t=1000$ generations, because
the dynamics converges well before that.

\subsection*{Results}\label{WFresults}
\begin{figure}[ht]
\centering
\includegraphics[width=17.5cm]{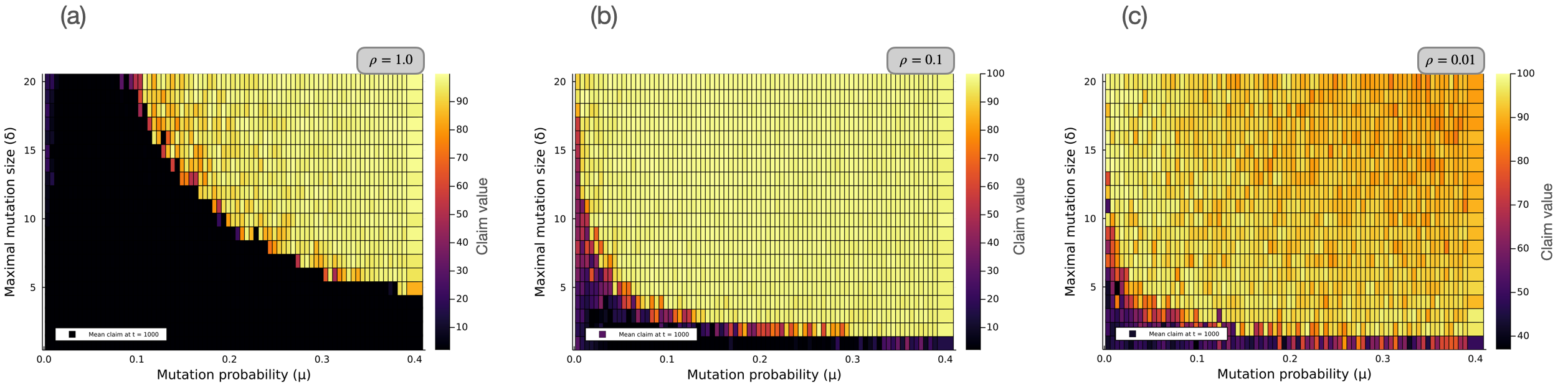}
\caption{\textbf{Wright-Fisher model
    with selection and mutation for the Traveler's Dilemma.} The graphs show the mean claim of a population of size $N=100$ after $t=1000$ time steps, for $R=2$ and different selection intensity values: \textbf{a)} $\rho=1.0$ \textbf{b)} $\rho=0.1$ \textbf{c)} $\rho=0.01$. In general, a large mutation probability ($\mu$) and a large maximal mutation size ($\delta$) allow the convergence to high claims, given that these quantities are proportional to the diversity present in the population. For the strong selection regime, the system tends to converge to low claims, because a high value of $\rho$ reduces the effective population diversity by magnifying even small payoff differences. For very low $\rho$ values, the level of stochasticity increases, given that the selection intensity parameter modulates the impact of fitness differences in the game dynamics.}
\label{fig:WF}
\end{figure}

For the Wright-Fisher model, a large mutation probability, $\mu$, and a large
maximal mutation size, $\delta$, enables the convergence to high claims (see
Fig. \ref{fig:WF}). The reason for this is that mutation introduces diversity
in the population, larger  mutation rates and sizes lead to greater diversity,
and in particular, to the appearance of  high claim individuals in the
population. When these ``suboptimal actions"  play against each other, a higher payoff results,
both collectively and individually.\\

In figure \ref{fig:WF}, it can be observed that for large $\rho$, the
system converges in a higher proportion to a low claim. The reason for this is that a high
value of $\rho$ reduces the population's diversity, as by the exponential
payoff-to-fitness mapping of equation \ref{eq:WF_Fitness}; given that $\rho$ is the
amplification level of fitness differences. Therefore, a large value of
$\rho$ magnifies even small fitness differences, that is, amplifies the differences between sampling weights.
Consequently, for high $\rho$ values, a global maximisation process can only
appear if the mutation rate and size are sufficiently large. Otherwise, the maximisation process will be constrained to a local level, which produces the convergence to the Nash equilibrium via the \textit{stairway to hell}.
It can also be noted that when $\rho$ is low, the level of stochasticity is large, which means that noise dominates the game dynamics, instead of the payoff differences.\\

Overall, the Wright-Fisher model helps to further understand diversity as the underlying mechanism that allows the convergence to a cooperative high claim equilibrium. Particularly, it shows how the local maximisation process can be escaped thanks to diversity in a population based model including genetic drift, selection and mutation.

\pagebreak

\section*{Introspection Dynamics}\label{intro}

\subsection*{Model}
Introspection dynamics is a learning model used to describe adaptive behaviour via introvert reasoning
\cite{hauser:Nature:2019,couto:NJP:2022}. The basic idea is that a player, after considering her payoff
in an encounter with an opponent, checks whether a randomly chosen alternative
action -sampled from the uniform distribution on the $[L,U]$ interval-, would have given her a better payoff against a given action chosen by her opponent. This process is called the introspection. A new action then is adopted with a probability depending on the payoff difference. Thus, when the payoff of the realised action is $\pi$,
while the alternative action would have yielded $\tilde{\pi}$, the
probability of adopting the latter action is computed using the Fermi function \cite{blume:GEB:1993, szabo:PRE:1998, hauert:AJP:2005}

\begin{equation}
\label{eq:Fermi}
p = \frac{1}{1+e^{- \beta(\tilde{\pi} - \pi)}}
\end{equation}
with a selection intensity coefficient $\beta$.
For weak selection ($\beta \rightarrow 0$), payoffs are not important in the
learning process as any alternative action is randomly adopted. For strong
selection ($\beta \rightarrow \infty$), players exclusively adopt the
alternative action if their payoff matches at least the realised payoff
\cite{couto:NJP:2022}. Therefore, $\beta$ determines the magnitude of the impact
that payoffs have on the game dynamics.
As we shall see, $\beta$ and the parameter $\rho$ of the Wright-Fisher model have different effects on the game dynamics, even though they share the same interpretation.\\

In this learning model, there are only two players that from time to time perform the introspection process, i.e.  consider the adoption of a counterfactual action given that their opponent's action is fixed. For the Traveler's Dilemma, the model is used in the context of a one-shot game, where players adjust their behaviour using introspection.\\

While the proposed model may be stylised and coarse, the principle is not psychologically implausible, as humans can both evaluate putative alternatives and perform a coarse choice when the number of details in the decision process, here the number of possible claims, becomes too large.\\

The model thus assumes a rather simple cognitive mechanism, but as we shall
see, this can lead to the emergence of cooperation for a wide range of
selection coefficients ($\beta$). This is a novel approach and result in the
context of the Traveler's Dilemma, where noise and errors
have been previously proposed as the cause for this behaviour.

\subsection*{Results}
\begin{figure}[ht]
\centering
\includegraphics[width=17.5cm]{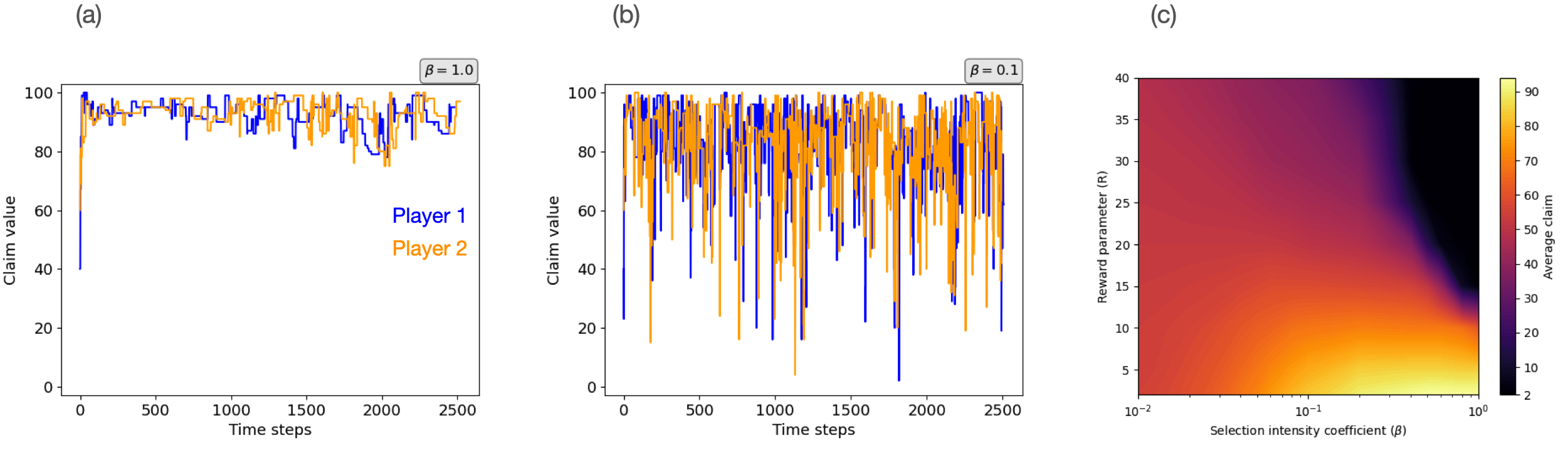}
\caption{\textbf{Introspection dynamics for the Traveler's Dilemma.}
Dynamics of the system for strong, \textbf{a)} $\beta=1.0$, and weak selection intensity, \textbf{b)} $\beta=0.1$), both for $R=2$. For the strong selection regime, the system converges to large claims thanks to the global optimisation process. In particular, the system oscillates in the high claim region, given that at a local level the system does not have a maximum. On the other hand, for weak selection the system becomes increasingly stochastic, until it fluctuates across all the action space.\\
\textbf{c)} The stationary distribution of actions allows the analytical calculation of the average claim for different reward parameter (R) and selection intensity ($\beta$) values. Strong selection and a low reward are needed for convergence to high claims.}
\label{fig:Intr_Dyn}
\end{figure}

For the introspection dynamics, the system converges to large claims in the strong selection regime for low values of R (see Figure \ref{fig:Intr_Dyn}.a).
The dynamics, however, does not converge to a specific value, but instead
oscillates in the high claim region. This is expected from the way the
optimisation process operates. First, the global optimisation is performed,
then the model shifts to a local maximisation. Given that at a local level the
system does not have a maximum -- due to the \textit{stairway to hell} --, the system
exhibits oscillations in the large claim region. For the weak selection
regime, as expected, the system becomes increasingly stochastic, until it comes to fluctuate across the whole action space (see Figure \ref{fig:Intr_Dyn}.b).\\

The stationary distribution of actions, derived using the main analytical result of Couto, Giaimo \& Hilbe. (2022) \cite{couto:NJP:2022}, allows the calculation of the average claim for different reward parameter (R) and selection intensity coefficient ($\beta$) values. As shown in figure \ref{fig:Intr_Dyn}.c, strong selection and a low reward value are requirements for the convergence to high claims. This result is in agreement with economic experiments findings, where individuals choose high claims when the reward parameter is low. Such decision is purely motivated by payoff maximisation, as it is in the case of strong selection, and not by mistakes or noise in the decision making process.\\

For the convergence to large claims, of course some diversity is needed, that
is, the presence of a certain fraction of large claim players or a sufficiently
high probability that agents select large claims. The random
adoption of an action that outperforms the current one in the introspection dynamics generates and
maintains such diversity.\\

Thus, the parameters $\rho$ and $\beta$ cause an overall different effect in the system's dynamics.
For the Wright-Fisher scenario, high values of $\rho$ prevent a global maximisation process by reducing the effective diversity of the population. On the other hand, $\beta$ does not have any effect on the diversity for the introspection dynamics, because the alternative random strategy is always chosen from a uniform distribution regardless of the magnitude of $\beta$.
However, it should be noted that $\rho$ and $\beta$ share the same
interpretation, as they are both parameters that modulate the impact of payoffs on the game dynamics. In particular, when $\rho$ or $\beta$ have low values, the level of stochasticity is large, which means that noise dominates the game dynamics, instead of the payoff differences.\\

Finally, we should note that, in contrast to our findings for the Traveler's Dilemma, cooperation does not emerge
in the Prisoner's Dilemma when played using introspection dynamics.
There, both players learn to defect at all times, that is, play Nash, in the limit
of strong selection \cite{couto:NJP:2022}. Therefore, diversity is only effective on promoting cooperation when the structure of the game allows it. As mentioned, the Traveler's Dilemma is composed on a local and a global game, where jumping from the local to the global level maximisation is required to reach a cooperative equilibrium.

\pagebreak

\section*{Comparison to other models}\label{other}
Previous models propose different methods to explain the convergence to high
claims in the Traveler's Dilemma. From our perspective, however, they operate
on the same underlying mechanism to allow for cooperation, namely diversity
enabling a global maximisation to dominate over a local optimisation and thereby
preventing the \textit{stairway to hell}.\\

The two main models are Goeree \& Holt. (1999) \cite{goeree:PNAS:1999} and Manapat et al. (2012) \cite{manapat:JTB:2012}. The first is a dynamic learning model with a logit rule, for which a player's decision probabilities are proportional to an exponential function of expected payoffs. In general, Goeree and Holt propose in their paper \cite{goeree:PNAS:1999} that ``noisy learning" allows for players to converge to high claims. This means that when players have the possibility to adopt non-Nash or ``suboptimal" strategies, due to a ``noisy learning process", the system can converge to high values. In our perspective, this is equivalent to having diversity in the game such that a global maximisation process occurs.\\

On the other hand, the stochastic evolutionary model formulated by Manapat et
al. (2012) \cite{manapat:JTB:2012} employs mutations to reach high claims. Overall, the results we
obtained using the Wright-Fisher model are in agreement with their results, as
they also find it necessary to include mutations in their model to reach large
values. Also, our and their approach exhibit the same behaviour in terms of the
selection intensity coefficient. Nonetheless, in Manapat et al. diversity is not identified as the general underlying mechanism for the convergence to high claims. We have explained in the Wright-Fisher model section how diversity enables the jump towards cooperation in a population, when the Traveler's Dilemma is analysed using stochastic evolutionary dynamics.\\

\section*{Discussion}
We have shown that diversity is the underlying mechanism that enables cooperation to emerge in a game like the Traveler's Dilemma. This mechanism avoids the paradox of classical game theory where supposedly rational players inevitably end up in a Nash equilibrium, which neither maximises the individual nor the collective payoff. As already pointed out by Kaushik Basu, who formulated the Traveler's Dilemma \cite{basu:AEA:1994}, and as confirmed by economic experiments \cite{capra:AER:1999}, such behaviour seems implausible for human players when faced with this game.\\

To analyse the dilemma, we proposed its decomposition into a local and a global game. Locally, the game structure is of a Prisoner's Dilemma, while globally it is a coordination game. In consequence, for a cooperative high claim equilibrium to be reached, it is necessary to jump from the local to the global payoff maximisation. In order to explore this idea, we used a dynamical approach, including population based models and a learning model. The replicator-mutator equation and the Wright-Fisher model for the former, and introspection dynamics for the latter.
Comparing the results, allowed us to identify a general mechanism that may lead to the convergence to strategies far from the Nash equilibrium that leave the players much better off. Such mechanism that favours cooperation is diversity. Here, diversity, understood as the presence of ``suboptimal strategies'', allows the jump from the local to the global level maximisation. Overall this can lead to cooperation because by considering ``suboptimal strategies" that play against each other it is possible to reach higher payoffs, both collectively and individually\\

This perspective also allowed us to identify diversity as the main mechanism of previous models used to analysed the Traveler's Dilemma \cite{goeree:PNAS:1999,manapat:JTB:2012}, although these models operate through very different formulations.\\

Our approach may also help to better understand how humans play a game like
the Traveler's Dilemma \cite{capra:AER:1999}, typically settling on high claims
instead of going to the Nash equilibrium. While the introspection dynamics is
admittedly rather stylised and coarse, it is not completely implausible as humans also tend to first make a coarse decision when faced with a large number of options. And in a game like the Traveler's Dilemma, a belief that opponents play ``suboptimal" actions may become self-confirming when it leads to higher payoffs. From our perspective, it suffices that the probability to encounter a suboptimal high
claim action to be sufficiently high. This is ensured by diversity, as previously explained.\\

Overall, the mechanism of diversity proposed here benefits players both at the collective and at the individual level. Therefore, exploring other games that share the same characteristic of having conflicting interests between a local and a global payoff maximisation might be helpful to understand and induce cooperation in scenarios like price wars. A situation where undercutting the opponent is tempting but reaching a cooperative equilibrium would be beneficial for everybody.

\section*{Code availability}

The code for the simulations and numerical solutions is available at  \url{https://github.com/MA-Ramirez/Travelers_Dilemma_Code.git}\\

\bibliography{refs}

\section*{Acknowledgements}

We would like to thank Christian Hilbe and Marta Couto for valuable feedback and discussions.

\section*{Author contributions statement}
M.A.R. conceived the study, J.J. and M.S. contributed. M.A.R. and M.S. performed the computer simulations.\\ M.A.R. computed the numerical solutions. All authors analysed and discussed the results. M.A.R. and J.J. wrote the manuscript.  All authors reviewed the manuscript.

\section*{Additional information}

\textbf{Competing interests}

The authors declare no competing interests.

\end{document}